\RequirePackage{lineno}
\documentclass[aps,prl,a4paper,preprintnumbers,twocolumn,floatfix,showpacs]{revtex4}
\usepackage{graphicx,graphics,epsfig,epic,eepic}    
\usepackage{color}
\usepackage{subfigure}
\usepackage{dcolumn}    
\usepackage{amsmath,amsfonts,amssymb}
\usepackage{mathrsfs}   	
\usepackage{bm}              
\usepackage{psfrag}
\usepackage{units}      	
\usepackage[colorlinks=true,linkcolor=blue,citecolor=blue]{hyperref}
\usepackage{enumerate}
\usepackage[all]{xy}
\begin{document}
\title{Reducing Spacetime to Binary Information}
\author{Silke Weinfurtner$^{1}$, Gemma De las Cuevas$^{2,3,4}$,  Miguel Angel Martin-Delgado$^{5}$ and Hans J.~Briegel$^{3,4}$}
\affiliation{
$^1$SISSA, Via Bonomea 265, 34136, Trieste, Italy {\rm and} INFN, Sezione di Trieste \\
$^2$Max-Planck-Institut f\"ur Quantenoptik, Hans-Kopfermann-Str. 1, D-85748, Garching, Germany \\
$^3$Institut f{\"u}r Theoretische Physik, Universit{\"a}t Innsbruck, Technikerstra{\ss}e 25, A-6020 Innsbruck, Austria\\ 
$^4$Institut f\"ur Quantenoptik und Quanteninformation der \"Osterreichischen Akademie der Wissenschaften, A-6020 Innsbruck, Austria\\
$^5$Departamento de F\'{\i}sica Te\'orica I, Universidad Complutense, 28040 Madrid, Spain}
\begin{abstract}
We present a new description of discrete space-time in 1+1 dimensions in terms of a set of 
elementary geometrical units that represent its independent classical degrees of freedom.
This is achieved by means of a binary encoding that is ergodic
in the class of space-time manifolds respecting coordinate invariance of general relativity.
Space-time  fluctuations can be represented in a classical lattice gas model whose Boltzmann weights 
are constructed with the discretized form of the Einstein-Hilbert action.
Within this framework, it is possible  to compute  basic quantities such as
the Ricci curvature tensor and the Einstein equations, and  to evaluate the path integral of discrete gravity. 
The description as a lattice gas model also  provides a  novel way of quantization
and, at the same time,  to quantum simulation of fluctuating space-time.
\end{abstract}
\maketitle

A central task in defining a discrete version of spacetime is to identify the relevant degrees of freedom. Additionally, any theory of discrete spacetime must take into account coordinate invariance, which is a fundamental property of General Relativity (GR). This renders the identification of fundamental degrees of freedom non-trivial. 
Solving this problem corresponds to fixing the gauge of the coordinate invariance symmetry of GR at the discrete level.

Here we address this question and present a binary description of space-time in 1+1 dimensions. To do so, we borrow the discretization method of causal dynamical triangulation (CDT) \cite{Ambjorn:2000fk}. Following this route leads to a formulation of spacetimes in terms of a statistical mechanical model, which we identify with a lattice gas model \cite{St87}. Fluctuations of spacetime are thereby  represented by different states of the lattice gas, and the Boltzmann factor for this statistical model is given by the discretized action of general relativity \cite{Regge:1961px}. 

The central idea of this work is to replace the dynamical geometry of space-time with a fixed physical lattice with binary degrees of freedom at its vertices. This construction allows us to digitalize the geometrical information content of space-times. 
More precisely, we show how to construct a foliated triangulation $T$ from a bit array $\lambda$ by using `forks', and conversely how to construct a bit array $\lambda$ from a foliated triangulation $T$. For the latter construction, one first maps $T$ to its dual $T^{*}$, $T^{*}$ to an integer string $S$, and $S$ to $\lambda$, as indicated in the following commuting diagram (the details of which will be explained below).
\begin{equation} 
\begin{aligned}
\xymatrix@!C{
\small{\textrm{Triangulation } T} \ar[d]    \:\: & \:\: \ar[l]^{\textrm{forks}} \small{\textrm{Bit array } \lambda} \\
\small{\textrm{Dual triangulation } T^{*}} \ar[r]  \:\:& \:\: \small{\textrm{Integer string } S}   \ar[u]  
}
\label{diagram1}
\end{aligned}
\end{equation}
In addition, we will translate the Pachner moves to operations on the integer string encoding. (The Pachner moves are transformations in the set of triangulations which are ergodic, that is, any two triangulations are related by a finite sequence of Pachner moves \cite{Pachner:1991lr}.) Finally, we will use our binary encoding to formulate meaningful quantities of discrete gravity in 1+1 dimensions in the natural language of information processing.

In order to establish these results,  let us first recall some basic properties of CDT~\cite{Ambjorn:2000fk}. 
In CDT the continuous manifold of space-time is approximated by a piecewise linear manifold \cite{David:1992jw}, where the edge lengths of the simplices are assumed to be constant (space-like edges have length $l^2_\mathrm{edge}$ and time-like edges $-\alpha l^2_\mathrm{edge}$, we shall henceforth assume $\alpha=-1$, which is one of the points in the `Euclidean sector' that gives rise to an interesting new phase \cite{Ambjorn:2004qm}). 
Additionally, only manifolds that obey a global proper discrete time are considered, i.e.manifolds with a discrete global time foliation. On the simplicial manifold one can define curvature, an action, and other quantities \cite{David:1992jw}.

\begin{figure}[t]
\includegraphics[width=0.49\textwidth]{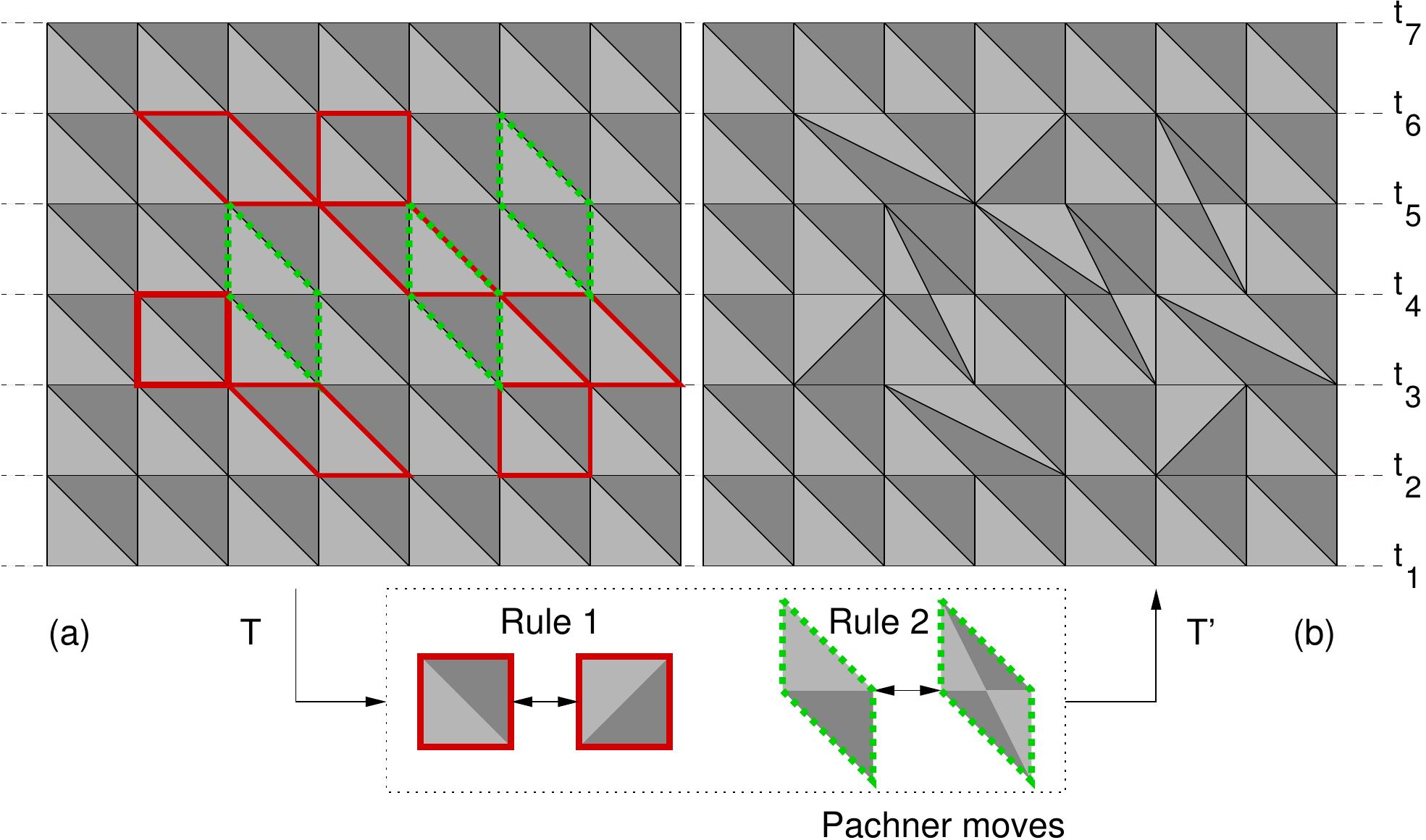}
\caption{
CDT considers only triangulations that obey a global time foliation. 
(a) A foliated triangulation corresponding to a flat geometry, since the coordination number of every vertex is 6, excluding the boundaries (see Eq.~\eqref{Equ:Ricci}). 
Any other foliated triangulation, such as (b), can be obtained by applying a sequence of the Pachner moves Rule 1 and Rule 2.}  
\label{FIG_1_new} 
\end{figure}

We shall here focus on the case of 1+1 dimensions, in which the configuration space $\mathcal{T}_t$ is formed by all foliated triangulations $T$ with with $t$ discrete proper-time steps (see Fig.~\ref{FIG_1_new}). 
Additionally, we restrict ourselves to simplicial manifolds with a fixed topology, such as $S^{1}\times S^{1}$ (periodic boundary conditions (pbc) in time and space), or  $[0,1]\times S^{1}$ (open boundary conditions in time and pbc in space). We consider only connected simplicial manifolds; this implies, in particular, that there is at least one simplex per spatial slice. 
The topology of the simplicial manifold is characterized by the Euler characteristic $\chi := N_0-N_1+N_2$, where $N_{0}, N_{1}, N_{2}$ is the number of vertices, edges and faces of $T$, respectively. Note that in 2 dimensions $\chi$ is related to the curvature via the Gauss--Bonnet theorem.

In this setting, the Pachner moves consist of two operations (called Rule 1 and Rule 2; see Fig.~\ref{FIG_1_new}).
The curvature $R_{i}$ is evaluated at every vertex $i$:
\begin{equation} \label{Equ:Ricci}
R_i = \pi \frac{6-c_i}{c_i}
\end{equation}
where $c_{i}$ is the coordination number (i.e. number of adjacent vertices) of $i$. 
The discretized Einstein--Hilbert action takes the form
\begin{equation} \label{Eqn:2Daction}
S(T) = \gamma \chi -\kappa N_2,
\end{equation}
where $\gamma$ and $\kappa$ are coupling constants related to Newton's constant $G_N$ and the effective cosmological constant $\Lambda$ in the continuum, respectively \cite{Ambjorn:2000fk}.
The quantization is carried out by means of a path integral formulation of the action \eqref{Eqn:2Daction}, which (for a specific topology such as $[0,1]\times S^{1}$) takes the form
\begin{equation}
\label{eq:Z}
Z=\sum_{T\in \mathcal{T}_{t}([0,1]\times S^{1})} \frac{1}{C({T})} e^{-S(T)} .
\end{equation}
Here $C({T})$ is the order of the automorphism group of $T$, and is the remnant of coordinate invariance of GR in the discrete theory.

Considerable progress has been made over the last few years ---in terms of analytic in $1+1$~\cite{Ambjorn:1998xu} and numerical studies in $2+1$~\cite{Benedetti:2007pp} and $3+1$~\cite{Ambjorn:2004qm} dimensions--- to investigate the continuum limit of CDT. 
Interesting connections with other continuum quantum gravity approaches have also been established~\cite{Horava:2009rt,Sotiriou:2011fk,Reuter:2011ah}. A connection between foliated triangulations in terms of random walks was established in \cite{Di00b}. Finally, let us mention that other discrete models of quantum gravity have been proposed, e.g. \cite{Bombelli:1987aa,Ko08}.

\begin{figure}
\includegraphics[width=0.49\textwidth]{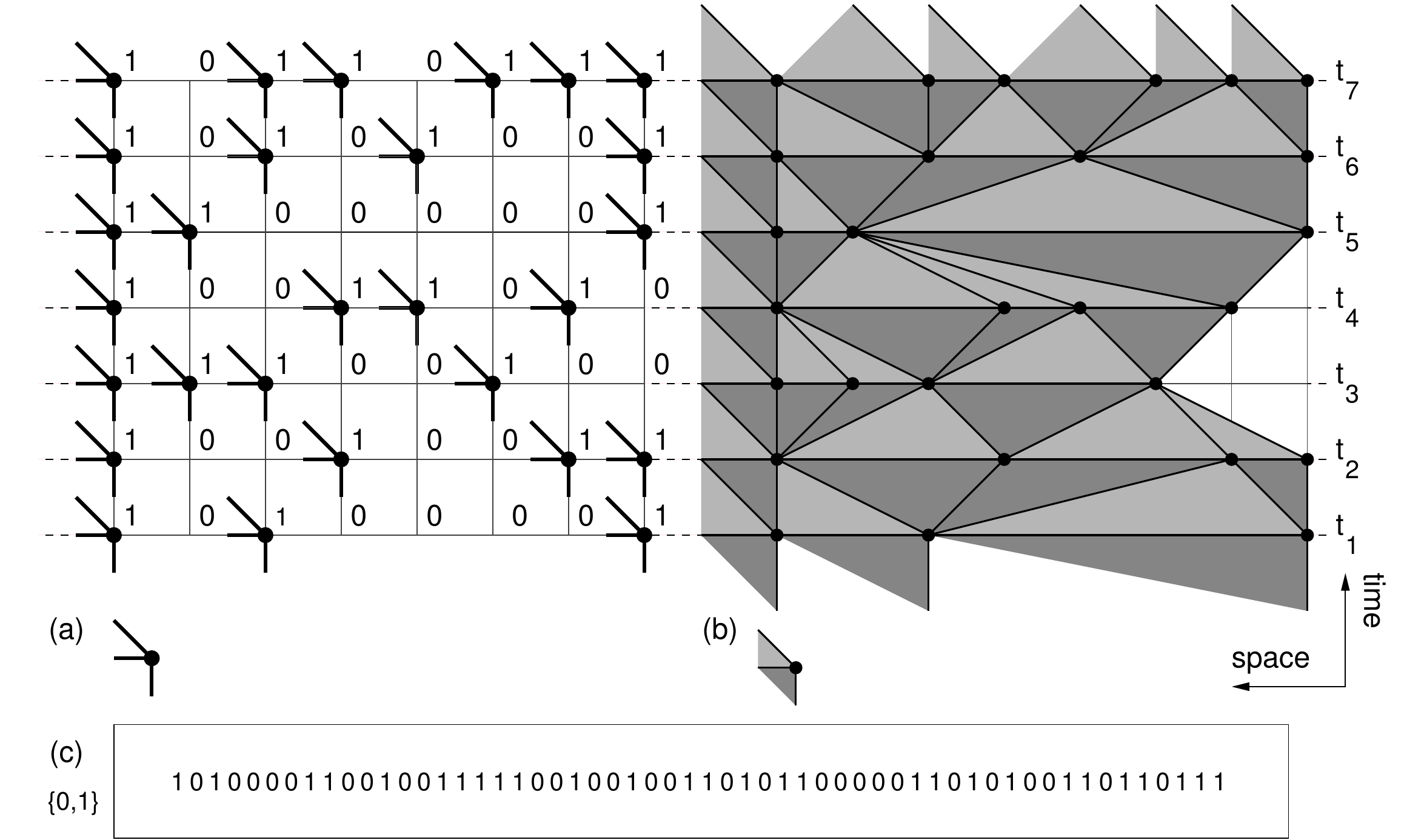}
\caption{\label{FIG_2_new}\emph{From bit arrays to triangulations.} 
The bit array (a) represents the foliated triangulation (b). Each bit stands for the presence / absence of a fork, which consists of one vertex, one space-like edge and two time-like edges.
While the bits are fixed on a two-dimensional grid, their configuration determine the geometry, similar to a  lattice gas model. 
(For illustration issues we have chosen non-periodic boundary conditions in both time and space.) To store the triangulation (b) on a double precision floating-point format (64-bit) machine a single data unit is required (c).}
\end{figure}

\emph{From bit arrays to triangulations.}
We now show how a bit array $\lambda$ encodes a foliated triangulation $T$, as indicated in Diagram~\eqref{diagram1}. 
The key observation is that every foliated triangulation $T$ is built entirely (except for the boundaries) out of certain building blocks that we call 'forks'. 
The idea is the following: while, obviously, the building block of a general triangulation is the triangle, the basic unit of a foliated triangulation is a pair of triangles that share a space-like edge. 
We identify the fork with this unit; more precisely, each fork consists on 1 `center', 3 legs and 2 faces (see little diagram on bottom of Fig.~\ref{FIG_2_new}(b)).
Thus, we can describe a foliated triangulation by `comparing' it to a reference lattice and specifying what forks are present and what forks are absent.
This renders a description which is in spirit similar to that of a lattice gas model, where the description of a fluid, with molecules absent or present, is mapped to the description of a magnet with two-level spins on a fixed lattice \cite{St87}. 

Let us be more precise. Consider a 2D square lattice where a binary variable $\lambda_{nm}$ is associated with every vertex $(n,m)$ of the lattice, with $1\leq n\leq N$ and $1\leq m\leq M$ (see Fig.~\ref{FIG_2_new}(a)). This forms a bit array $\lambda$. To transform this bit array to a foliated triangulation, we put a fork on every site $(n,m)$ it $\lambda_{nm}=1$, and no fork if $\lambda_{nm}=0$. This collection of forks defines a graph $T=(V,E)$ in a natural way: each center of a fork defines a vertex $v\in V$, and its legs become edges $e\in E$ of $T$. The space-like edge and the time-like edge pointing upwards connect to the first vertex to the left, and the time-like edge pointing downwards connects to the first vertex directly below or to the left (see Fig.~\ref{FIG_2_new}). 
This can be formally described as follows. Let $p$ be a rank 3 tensor with components
\begin{equation} \label{Eq:findindex}
p_{n,m,m'}:= \lambda_{nm'}  \times [(m-m' \pmod M) +1] \, \, .
\end{equation} 
Then the vertex at site $(n,m)$ is adjacent to the vertices at sites $(n+1,a)$, $(n,b)$, $(n-1,c)$ where $a$, $b$, $c$ are defined by the inequalities
\begin{equation}  \label{Equ:ConnectionRules}
\begin{array}{cl}
0 < p_{n+1,m-1,a} < p_{n+1,m-1,m'} \quad  &\forall m'\in [M], \: m'\neq a\\
0 < p_{n,m-1,b} < p_{n,m-1,m'} \quad & \forall m'\in [M], \: m' \neq b\\
0 < p_{n-1,m,c} < p_{n-1,m,m'} \quad & \forall m'\in [M], \: m' \neq c,
\end{array}
\end{equation}
where $[M]=\{1,2,\ldots, M\}$.

\begin{figure*}
\includegraphics[width=0.85\textwidth]{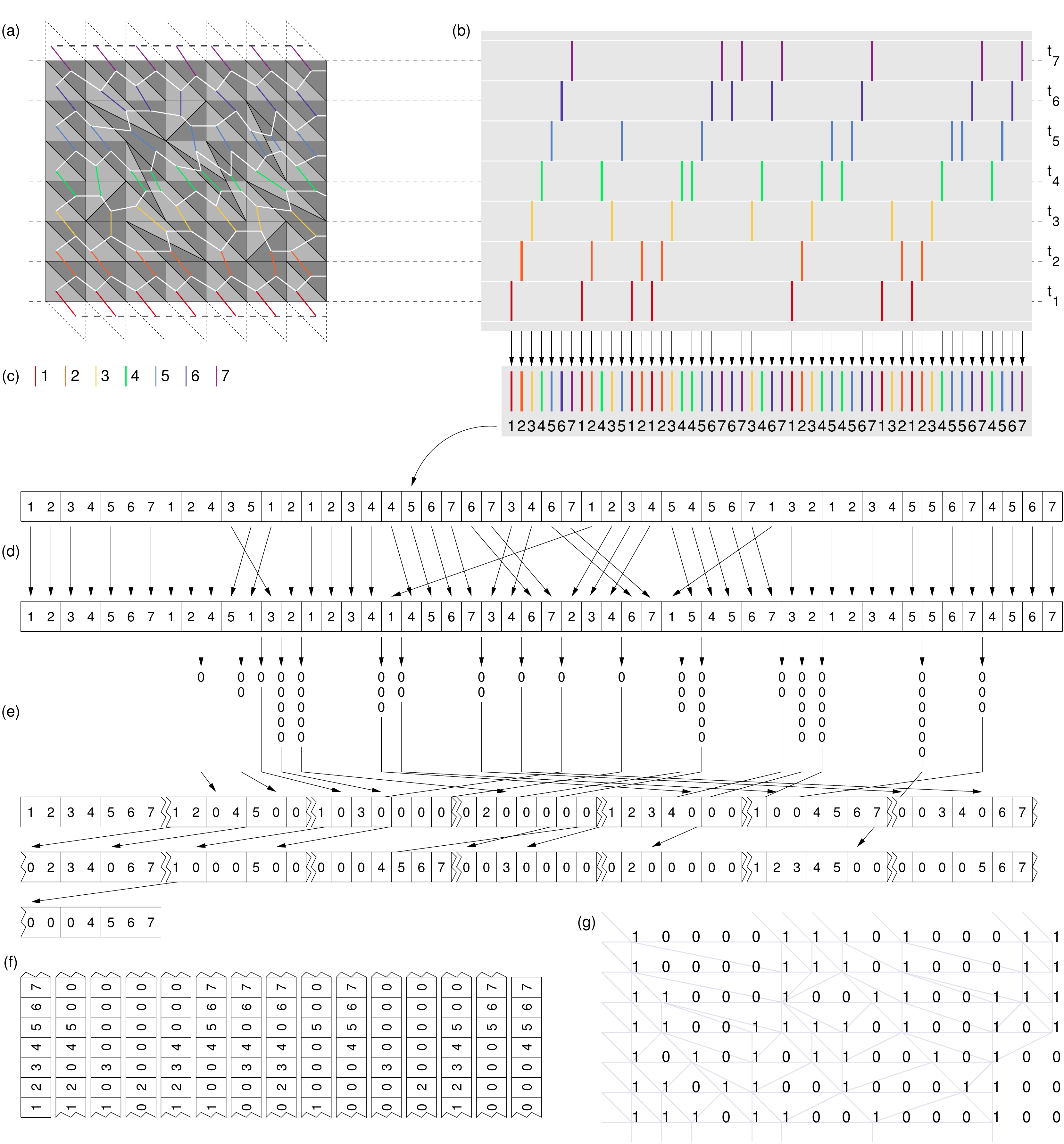}
\caption{\label{FIG_3_new} \emph{From triangulations to bit arrays.} 
The triangulation $T$ of panel (a) (which is the same as in Fig.~\ref{FIG_1_new}(b)) is mapped to the bit array of panel (g). (a) We first construct the dual of $T$ (a), and redraw it to give it the appearance of a brick wall (b), where every vertical (colored) line represents a dual fork (d-fork). The dual graph is mapped to a string $S$ (c). 
Using operations \eqref{i}-\eqref{iii} we reorder $S$ in successive incomplete integer-sequences (d), and extend the string by adding zeros where the integer-sequence is incomplete (e). Finally, we map this extended string to the bit array with Eq.~\eqref{Eq:Mapping} (f,g). (Notice that we consider here open boundary conditions on the spatial slices.)
\vspace{-0.5cm}}
\end{figure*}

\emph{From triangulations to bit arrays.} 
We now show how to map a foliated triangulation to a bit array. The recipe consists of six steps.  
First, construct the dual graph of $T$, denoted $T^{*}$, whose basic building block is the d-fork (short for dual fork). This consists of one time-like edge and two space-like edges ``pointing to the right'' (see Fig.~\ref{FIG_3_new}(a,b)). 
Second, note that every d-fork crosses one space-like edge of $T$. We use this fact to attach the label $S_{i}\in \{ 1,2, \dots, N \}$ to a fork $i$ if it crosses a space-like edge at time $t_{S_{i}}$ (see Fig.~\ref{FIG_3_new}(c), where the labels are also represented as colors). 
Third, record the labels of the d-forks that appear in $T^{*}$ from left to right and write them down in an integer string $S:= (S_1,S_2, \dots, S_F)$ (cf. Diagram \eqref{diagram1}). 
This string representation has certain symmetries. Two integer strings correspond to the same dual triangulation if they are related by a sequence of the following operations:
\begin{enumerate}[(i)]
\item \label{i}
commutations of contiguous integers if the integers differ by at least two, i.e.~$(\dots ,S_i, S_{i+1} , \dots)\sim (\dots, S_{i+1}, S_i , \dots)$ if $\vert S_i - S_{i+1} \vert \geq 2$;
 \vspace{-0.3cm}
\item \label{ii}
cyclic permutations, i.e.~$(S_1, S_2, \dots , S_F)\sim (S_F, S_1, S_2, \dots, S_{F-1})$; 
\vspace{-0.3cm}
\item \label{iii}
inversion operation, i.e.~$(S_1,S_2, \dots, S_F) \sim  (S_F, S_{F-1}, \dots, S_1)$. 
\end{enumerate}
Notice that \eqref{ii} and \eqref{iii} are only applicable for periodic boundary conditions on the spatial slices, while \eqref{i} is a degeneracy introduced by the integer string encoding.  

Fourth, define a string $S^\textrm{flat}:=(S^\mathrm{N},S^\mathrm{N},\dots,S^\mathrm{N})$, consisting of complete integer-sequences, $S^\mathrm{N}:=(1,2,\dots,N)$. 
Apply operation \eqref{i} to arrange any $S$ in successive integer-sequences, by allowing for incomplete sequences, e.g.~$(1,3,4,6)$ (see Fig.~\ref{FIG_3_new}(d)). 
Fifth, define the extended string $S^{\textrm{E}}$ by adding zeros to $S$ wherever there is a missing element, e.g.~$(1,0,3,0,0,6)$ (see Fig.~\ref{FIG_3_new}(e)). Generally,
\begin{equation} \label{Eq:ExtendingString} 
S^\mathrm{E}:=(
\underbrace{0\dots 0}_{\Gamma_1},S_1, 
\underbrace{0\dots 0}_{\Gamma_2},S_2,   
\dots, S_F,
\underbrace{0\dots 0}_{\Gamma_{F+1}})
\end{equation}
where $A:=(N,S_1,S_2,\dots,S_F,1)$, and $\Gamma$ is a vector with components $\Gamma_i:=A_{i+1}-A_{i}+N-1 \pmod N$, for $i \in \{1,2,\dots, F+1\}$.

Sixth, map $S^\mathrm{E}$ to $\lambda$ by arranging the entries of S in a two dimensional array of size $N\times M$ (going from bottom to top and left to right), and then replacing each positive entry by a 1. Formally, 
\begin{equation} \label{Eq:Mapping}
\lambda_{ij} =
 \Theta(S^\mathrm{E}_{j \,N-i+1}) \quad \mbox{for} \quad \left\{
\begin{array}{l}
1\le i \le N  \\ 
1\le j \le \frac{F^\mathrm{E}}{N}
\end{array} \right. ,
\end{equation}
where $\Theta(x)= 0$ if $x=0$ and $\Theta(x)=1$ for $x>0$ (see Fig.~\ref{FIG_3_new}(f)). The minimum lattice size necessary to encode all triangulations with at most to $F$ forks with $N$ spatial slice is given by $N\times\tilde{M}$, where $\tilde{M}=F-N+1$ (see Supplementary Information \cite{SM}).

Note that by construction the integer encoding represents a coordinate-free encoding of triangulations.  Consequently, operations \eqref{i}-\eqref{iii} introduce an equivalence class on the space of all integer strings, and there exists a straightforward algorithm to single out its representatives. 
This can be illustrated with the help of a simple example, evaluating all unique histories for triangulations with 3 spatial slices from 3 to 5 forks. Starting from all possible strings, one can apply operations \eqref{i}-\eqref{iii} to single out one representative $\tilde S$ of every equivalence class: 
$(1,2,3)$, $(1,2,3,3)$, $(1,2,2,3)$, $(1,2,3,2)$, $(1,1,2,3)$, $(1,2,3,3,3)$, $(1,2,2,3,3)$, $(1,2,3,2,3)$, $(1,2,3,3,2)$, $(1,2,2,2,3)$, $(1,2,2,3,2)$, $(1,2,3,2,3)$, $(1,1,2,3,3)$, $(1,1,2,2,3)$, $(1,1,2,3,2)$, $(1,2,1,2,3)$, and $(1,1,1,2,3)$. 
See Supplementary Information \cite{SM} for a more thorough discussion.

\emph{Applications of binary encoding.}
We now rephrase various quantities of CDT in binary language. 
Consider first the Ricci scalar (Eq.~\eqref{Equ:Ricci}). 
The coordination number of a vertex at site $(n,m)$ is 3 (because each fork has 3 edges) plus additional $\Delta_{nm}$ edges which depend on the surrounding bit array; explicitly:
\begin{eqnarray}
\Delta_{nm}=\sum_{j=1}^{F(d-m)}&&[ 
\lambda_{n+1,F(m+j-1)}\nonumber \\ 
&+& \lambda_{n+1,F(m+j)} 
+ \lambda_{n+1,F(m+j)}]
\end{eqnarray}
where $F(m'):=m' \pmod M$, and $d$ is the index of the first non-zero entry to the left of $m$ given by $0<p(n,m+1,d)<p(n,m+1,m')\: \forall m'\neq d$ (see Eqs.~\eqref{Eq:findindex}, \eqref{Equ:ConnectionRules}).
The Ricci scalar $R_{nm}$ is then given by 
\begin{equation}
R_{nm}= \pi  \, \lambda_{nm}\, \frac{3-\Delta_{nm}}{3+\Delta_{nm}}. 
\end{equation}

The action (Eq.~\eqref{Eqn:2Daction}) is also easily computed in the binary encoding.  
To compute the Euler characteristic, note that every fork is associated to one vertex, 3 edges and 2 faces, hence it does not contribute to the value of $\chi$. Thus only forks relative to the boundary (that is, those which are placed at the boundary, either in the space or time dimension) contribute to $\chi$. For instance, for a torus (topology $S^{1}\times S^{1}$), $\chi=0$. 
The volume is then given by twice the number of forks, $N_{2}=2F$. 
This, together with $C(T)$, which can be evaluated exactly for 1+1 dimensional triangulations \cite{Di06}, suffices to evaluate the discretized path integral (Eq.~\eqref{eq:Z}).

Finally, the Pachner moves can also be expressed in the integer encoding (see Supplementary Information \cite{SM}). This allows us to show that the Pachner moves in 1+1 dimensions are ergodic, i.e~they generate all simplicial triangulations.

\emph{Conclusion and Outlook}. 
We have presented a binary encoding of discrete space-time in 1+1 dimensions. To this end, we have borrowed the discretization method of CDT and shown how to compute various quantities of interest of this theory in the binary description. 
Our results enable us to express a classical theory of discrete space-time as a lattice gas model. This approach has several potential applications. 
To begin with, it provides a natural  framework for quantization.
Forks, the elementary geometrical unit, constitute independent classical degrees of freedom or 'normal modes' of discrete space-time.
In the quantized theory, the presence or absence of a fork will be interpreted as elementary excitations of a quantum lattice gas. These excitations will give rise to quantum fluctuations
and, more generally, to quantum states representing superpositions of different space-times.
Despite using a time foliation, our approach to quantizing space-time is conceptually different from CDT quantum gravity, which is based on a path integral formulation.

Discretization of space-time was introduced as a regularization tool to compute the path integral of general relativity.
If one considers discrete space-time as {\it real}, then forks could be seen as fundamental
constituents of space-time, namely as basic geometric structures that connect events.
A quantum formulation of this theory would naturally introduce quantum correlations in space-time.

Finally, from an experimental perspective, by implementing quantum lattice gas models in the way we have introduced them, one could realize quantum simulations of fluctuating space-time. Such quantum simulators, analog or digital, could be used  to simulate models of quantum gravity in modern quantum optics laboratories, e.g. using ultra-cold atoms in optical lattices or ion-traps \cite{We10}.

\appendix

We acknowledge support by the Spanish MICINN grant FIS2009-10061, the CAM research consortium QUITEMADS 2009-ESP-1594, the European Commission PICC: FP7 2007-2013, Grant No. 249958, the UCM-BS grant GICC-910758, and the Austrian Science Fund (FWF) through project F04012.
SW was supported by Marie Curie Actions -- Career Integration Grant (CIG); Project acronym: MULTI-QG-2011 and the FQXi Mini-grant ``Physics without borders''. SW would like to thank Matt Visser, Piyush Jain, and Thomas Sotiriou for enlightening comments and discussions. GDLC acknowledges support from the Alexander von Humboldt foundation.

\begin{widetext}
\newpage
\section{\Large Supplementary Information}

We present supporting material for our paper. The focus is on specific details of, and relationships between, the various encodings of triangulations: (I) we discuss the efficiency of the binary encoding, and derive the minimum lattice size necessary to encode all triangulations with at most to $F$ forks distributed over $N$ spatial slices; (II) we will translate the Pachner moves to operations on the integer string encoding. The Pachner moves are transformations in the set of triangulations which are ergodic, that is, any two triangulations are related by a finite sequence of Pachner moves. This connection can be used to verify the ergodicity of the Pachner moves in 1+1 dimensions; (III) we show how to utilize the integer string encoding and the corresponding symmetry operations introduced in the main paper to single out integer strings representing unique triangulations; and (IV) we compare the degeneracies arising from the binary and integer string encoding. 
\subsection{Lattice size in binary encoding}
The minimum lattice size necessary to encode all triangulations $T$ up to $F$ forks distributed on $N$ spatial slices is given by $N\times\tilde{M}$, where $\tilde{M}=F-N+1$. To show this, let us temporarily put aside the requirement that there is at least one fork on every spatial slice, equivalent to the requirement that $S$ contains every element in $\{1,2,\dots,N\}$ at least once. It is easy to see that $S=\{X,X, \dots,X\}$, where all elements $X\in \{1,2,\dots,N\}$ are the same, maximizes $\Gamma=(N-X,N-1,N-1,\dots,N-1,X-1)$, where $\Gamma$ is a vector of length $F+1$, such that $F^\mathrm{E}=F+(N-1)  F=F\,N$ and $\tilde{M}=F$. Consequently, by reintroducing the constraint to have at least one fork per spatial slice, the string $S$ that will result in the largest $\tilde{M}$, has as many elements of the same kind, that are $F-N+1$ d-forks on a single spatial slice $X$. It is best to study this case using the dual brick picture, where it is straightforward to see that d-forks, $X\pm n$, for $n\ge1$, can be moved freely in both directions by using operations (i)-(iii) (see Fig.~\ref{FIG:latticeSize}). Therefore any of the $N-1$ d-forks can be absorbed in a sequence that contains already one $X$, and thus the number of sequences necessary to encode \emph{all} possible triangulations with $F$ forks distributed on $N$ spatial slices is $F-N+1$. According to the mapping, see Eq.~(7) in the main text, every sequence in $S^\mathrm{E}$ corresponds to one row in the corresponding $\lambda$.

\begin{figure}[htb]
\includegraphics{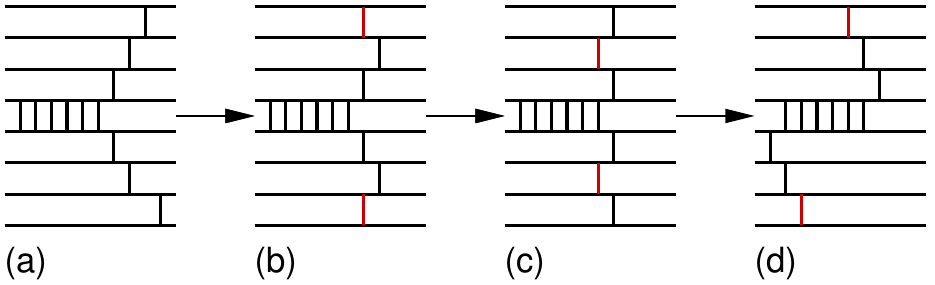}
\caption{Four different dual triangulating are represented (a-d). It is shown, that by successive application of operations (i)-(iii) of the main article acting on (a) one obtains (d). 
The the d-forks colored in red are moved with respect to the previous dual triangulations.}
\label{FIG:latticeSize}
\end{figure}

\section{Integer equivalent of Pachner moves}
\label{sec:integer-Pachner}
The Pachner moves are certain operations that when applied on a triangulation $T$ generate other triangulations $T'$. They are claimed to be ergodic, i.e. for any pair of triangulations, there always exists a sequence of Pachner moves that one can apply on one triangulation that yield the other. For foliated triangulations in 1+1 dimensions there are only two such moves (called Rule 1 and Rule 2, see Fig.~1 of the main text). Here we define the Pachner moves in the integer string description, that is, transformations on the integer strings which correspond to applying the Pachner move on the corresponding triangulation, as the following diagram illustrates.
\begin{equation}
\begin{aligned}
\xymatrixcolsep{3pc}\xymatrix{
T  \ar[d]^{\textrm{Pachner}} \ar[r] \quad & \quad T^{*}  \ar[r] \quad &\quad S \ar[d]^{\textrm{``integer Pachner''}} \\
T'  \ar[r] \quad & \quad (T')^{*} \ar[r] \quad &\quad S' \\}
\label{diagram2}
\end{aligned}
\end{equation}

\begin{figure*}
\includegraphics[width=1\textwidth]{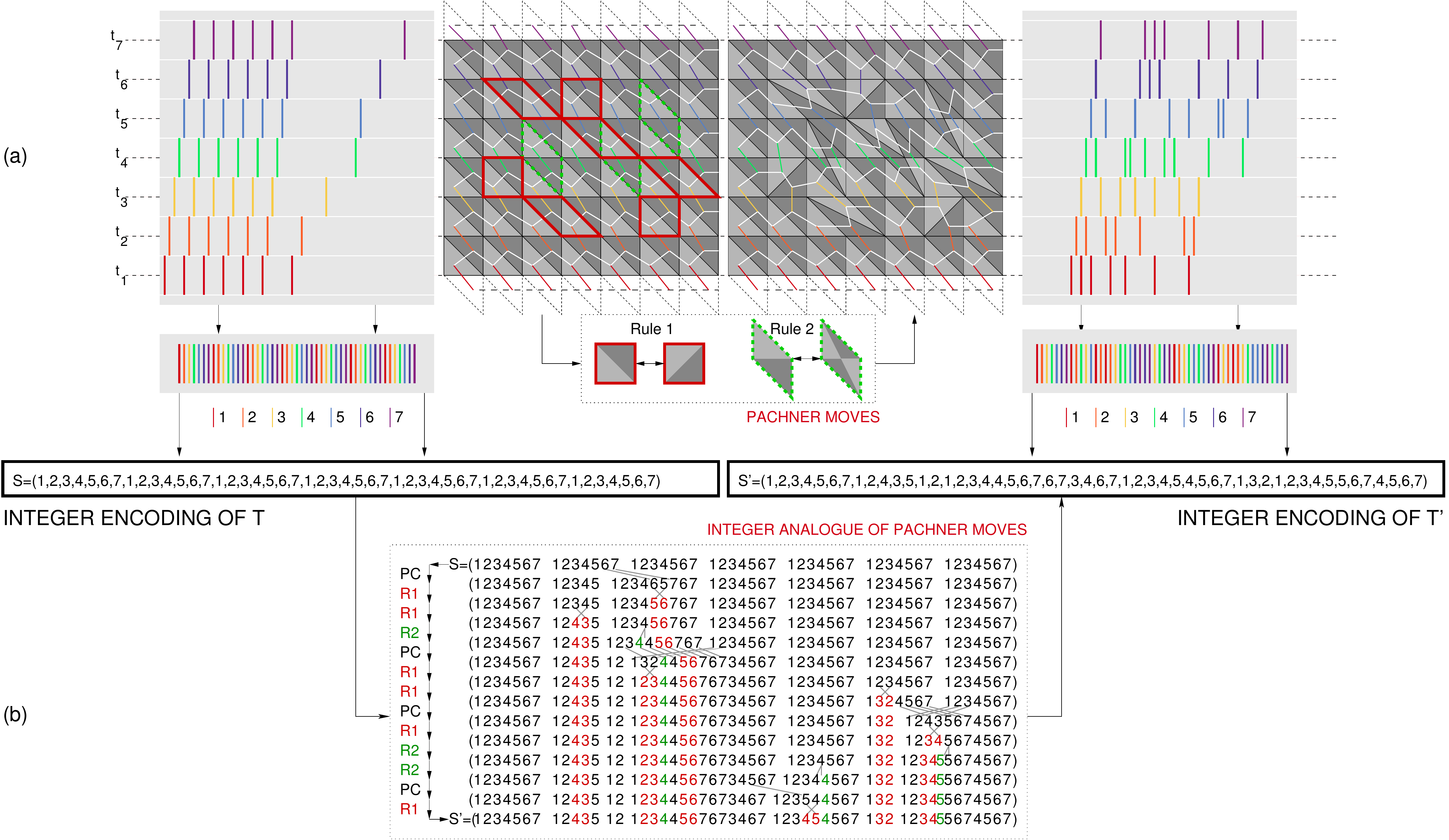}
\caption{\label{FIG_Pachner_dual_fork} 
We show explicitly how to apply successively R1 and R2  on $T$ to obtain $T'$ in both, the simplicial (a) and the integer string (b) encoding. In order to apply R1 it is sometimes necessary to use the pairwise commutation relations (PC), i.e.~$(\dots ,S_i, S_{i+1} , \dots)\sim (\dots, S_{i+1}, S_i , \dots)$ if $\vert S_i - S_{i+1} \vert \geq 2$. Notice that the integer strings $S$ and $S'$ correspond to the triangulations $T$ and $T'$ of Fig.~1 in the main text, respectively. } 
\end{figure*}

The action of Rule 1 (R1) is easily understood in the dual triangulation, where it simply exchanges the order of two neighboring consecutive d-forks. In the integer description it corresponds to the following operation
\begin{equation}
(\dots,S_{i},S_{i+1},\dots) \xrightarrow[]{\mbox{R1}} (\dots,S_{i+1},S_{i},\dots). 
\end{equation}
where $|S_{i}-S_{i+1}|=1$. 
Rule 2 (R2) has an equally simple equivalent, 
\begin{equation} 
(\dots,(n-1),n,(n+1)\dots)   \xrightarrow[]{\mbox{R2}}   
(\dots,(n-1),n,n,(n+1),\dots), 
\end{equation}
duplicating a d-fork and placing it right next to it. 
It is thus clear that R1 and R2 connect different equivalence classes in the string encoding. 
This demonstrates the ergodicity of the Pachner moves in 1+1 dimensions:
starting from the minimal string $S=(1,2,3,\dots,N-2,N-1,N)$, successive application of R1 and R2 generate all possible strings, and hence all possible triangulations.
Fig.~\ref{FIG_Pachner_dual_fork} shows the action of the ``integer Pachner moves'' on triangulations mentioned in the main text.

\section{Representatives of equivalent classes in all three encodings}
To evaluate the path integral of CDT in any dimensions analytically, one has to first single out triangulations which are representatives of their equivalence classes. We demonstrate how this procedure can be done in  integer string encoding. 
First, note that to encode all triangulations of size $V=2F$ distributed over $N$ spatial slices, one has to generate all strings containing at least $N$ forks and at most $F$ of them containing each element of $\{1 \dots N \}$ at least once. We then apply symmetry operations (i)-(iii) as illustrated in the following example.

Consider all triangulations containing up to 5 forks distributed over 3 spatial slices. We require at least one fork on every spatial slice; the smallest (in terms of 2-volume) triangulations contain 3 forks. The procedure to find the representatives for the equivalent classes is the following:
\begin{enumerate}
\item generate all possible integer strings of lengths $3$, $4$ and $5$, with elements in $S_i\in\{1,2,3\}$;
\item consider only strings that contain every element $\{1,2,3\}$ at least once (so that only connected triangulations are taken into account);
\item apply symmetry operations to single out unique triangulations\,/\,integer strings: if two integer strings are related by a sequence of the symmetry operations (i--ii) they are in the same equivalence class. One representative of each equivalence class give all unique histories\,/\,triangulations. It is best to demonstrate this at hand of an example. What are the representatives of triangulations containing 4 forks distributed over 3 spatial slices? From basic combinatorics we are left with nine valid integer string encodings: $S_1=(1,2,3,3)$, $S_2=(1,3,2,3)$, $S_3=(1,3,3,2)$, $S_4=(1,2,2,3)$, $S_5=(1,2,3,2)$, $S_6=(1,3,2,2)$, $S_7=(1,1,2,3)$, $S_8=(1,1,3,2)$, and $S_9=(1,2,1,3)$. Let us to begin with focus on the dual triangulations with one extra d-fork on the third row, i.e.~$S_1$, $S_2$ and $S_3$. Utilizing the symmetry operations we can show that $S_2=(1,3,2,3)=(3,1,2,3)=(1,2,3,3)=S_1$, and $S_3=(1,3,3,2)=(3,1,3,2)=(3,3,1,2)=(3,1,2,3)=(1,2,3,3)=S_1$, and consequently $S_2$ and $S_3$ are redundant, and $S_1$ is a representative. Similarly we get $S_6=(1,3,2,2)=(3,1,2,2)=(1,2,2,3)=S_4$, $S_8=(1,1,3,2)=(1,3,1,2)=(3,1,1,2)=(1,1,2,3)=S_7$, and $S_9=(1,2,3,1)=(1,1,2,3)=S_7$. Altogether we can for example single out the following representatives: $S_1$, $S_4$, $S_5$, and $S_7$. \end{enumerate}

Next we present the binary and the integer description for this set of triangulations explicitly.
For the binary description we need to consider arrays of size $N\times\tilde F-N+1=3 \times 3$ (see above).

\subsubsection{Representatives for 3 forks distributed over 3 spatial slices}

There is only one way to distribute 3 forks on 3 spatial slices, such that 
\begin{equation} 
\begin{array}{lclclcl}
\vspace{1mm}
\tilde S_1=(1,2,3) & \quad \rightarrow \quad &\tilde  S_1^\mathrm{E}=(1,2,3,\;0,0,0,\;0,0,0) &\quad \rightarrow \quad & \lambda(\tilde S_1^\mathrm{E})=\left(\begin{array}{ccc}
1 & 0 & 0 \\
1 & 0 & 0 \\
1 & 0 & 0
\end{array}\right), &\quad& $see FIG$.~\ref{FIG_representatives}(1) \\  
\end{array}
\end{equation}
 \subsubsection{Representatives for 4 forks distributed over 3 spatial slices}
After applying the symmetry operations we are left with four representatives:
\begin{equation} 
\begin{array}{lclclcl}
\vspace{1mm}
\tilde S_2=(1,2,3,3) & \quad \rightarrow \quad &\tilde  S_2^\mathrm{E}=(1,2,3,\;0,0,3,\;0,0,0) &\quad \rightarrow \quad & \lambda(\tilde S_2^\mathrm{E})=\left(\begin{array}{ccc}
1 & 1 & 0 \\
1 & 0 & 0 \\
1 & 0 & 0
\end{array}\right), &\quad& $see FIG$.~\ref{FIG_representatives}(2); \\  
\vspace{1mm}
\tilde S_3=(1,2,2,3) & \quad \rightarrow \quad &\tilde  S_3^\mathrm{E}=(1,2,0,\;0,2,3,\;0,0,0) &\quad \rightarrow \quad & \lambda(\tilde S_3^\mathrm{E})=\left(\begin{array}{ccc}
0 & 1 & 0 \\
1 & 1 & 0 \\
1 & 0 & 0
\end{array}\right), &\quad& $see FIG$.~\ref{FIG_representatives}(3); \\ 
\vspace{1mm}
\tilde S_4=(1,2,3,2) & \quad \rightarrow \quad &\tilde  S_4^\mathrm{E}=(1,2,3,\;0,2,0,\;0,0,0) &\quad \rightarrow \quad & \lambda(\tilde S_4^\mathrm{E})=\left(\begin{array}{ccc}
1 & 0 & 0 \\
1 & 1 & 0 \\
1 & 0 & 0
\end{array}\right), &\quad& $see FIG$.~\ref{FIG_representatives}(4) \\ 
\vspace{1mm}
\tilde S_5=(1,1,2,3) & \quad \rightarrow \quad &\tilde  S_5^\mathrm{E}=(1,2,3,\;1,0,0,\;0,0,0) &\quad \rightarrow \quad & \lambda(\tilde S_5^\mathrm{E})=\left(\begin{array}{ccc}
1 & 0 & 0 \\
1 & 0 & 0 \\
1 & 1 & 0
\end{array}\right), &\quad& $see FIG$.~\ref{FIG_representatives}(5).  
\end{array}
\end{equation}

 \subsubsection{Representatives for 5 forks distributed over 3 spatial slices}
After applying the symmetry operations we are left with 12 representatives:
\begin{equation} 
\begin{array}{lclclcl}
\vspace{1mm}
\tilde S_6=(1,2,3,3,3) & \quad \rightarrow \quad &\tilde  S_6^\mathrm{E}=(1,2,3,\;0,0,3,\;0,0,3) &\quad \rightarrow \quad & \lambda(\tilde S_2^\mathrm{E})=\left(\begin{array}{ccc}
1 & 1 & 1 \\
1 & 0 & 0 \\
1 & 0 & 0
\end{array}\right), &\quad& $see FIG$.~\ref{FIG_representatives}(6); \\  
\vspace{1mm}
\tilde S_7=(1,2,2,3,3) & \quad \rightarrow \quad &\tilde  S_7^\mathrm{E}=(1,2,0,\;0,2,3,\;0,0,3) &\quad \rightarrow \quad & \lambda(\tilde S_7^\mathrm{E})=\left(\begin{array}{ccc}
0 & 1 & 1 \\
1 & 1 & 0 \\
1 & 0 & 0
\end{array}\right), &\quad& $see FIG$.~\ref{FIG_representatives}(7); \\  
\vspace{1mm}
\tilde S_8=(1,2,3,2,3) & \quad \rightarrow \quad &\tilde  S_8^\mathrm{E}=(1,2,3,\;0,2,3,\;X,X,X) &\quad \rightarrow \quad & \lambda(\tilde S_8^\mathrm{E})=\left(\begin{array}{ccc}
1 & 1 & 0 \\
1 & 1 & 0 \\
1 & 0 & 0
\end{array}\right), &\quad& $see FIG$.~\ref{FIG_representatives}(8); \\  
\vspace{1mm}
\tilde S_9=(1,2,3,3,2) & \quad \rightarrow \quad &\tilde  S_9^\mathrm{E}=(1,2,3,\;0,0,3,\;0,2,0) &\quad \rightarrow \quad & \lambda(\tilde S_9^\mathrm{E})=\left(\begin{array}{ccc}
1 & 1 & 0 \\
1 & 0 & 1 \\
1 & 0 & 0
\end{array}\right), &\quad& $see FIG$.~\ref{FIG_representatives}(9); \\  
\vspace{1mm}
\tilde S_{10}=(1,2,2,2,3) & \quad \rightarrow \quad &\tilde  S_{10}^\mathrm{E}=(1,2,0,\;0,2,0,\;0,2,3) &\quad \rightarrow \quad & \lambda(\tilde S_{10}^\mathrm{E})=\left(\begin{array}{ccc}
0 & 0 & 1 \\
1 & 1 & 1 \\
1 & 0 & 0
\end{array}\right), &\quad& $see FIG$.~\ref{FIG_representatives}(10); \\  
\vspace{1mm}
\tilde S_{11}=(1,2,2,3,2) & \quad \rightarrow \quad &\tilde  S_{11}^\mathrm{E}=(1,2,0,\;0,2,3,\;0,2,0) &\quad \rightarrow \quad & \lambda(\tilde S_2^\mathrm{E})=\left(\begin{array}{ccc}
0 & 1 & 0 \\
1 & 1 & 1 \\
1 & 0 & 0
\end{array}\right), &\quad& $see FIG$.~\ref{FIG_representatives}(11); \\  
\vspace{1mm}
\tilde S_{12}=(1,2,3,2,3) & \quad \rightarrow \quad &\tilde  S_{12}^\mathrm{E}=(1,2,3,\;0,2,3,\;0,0,0) &\quad \rightarrow \quad & \lambda(\tilde S_{12}^\mathrm{E})=\left(\begin{array}{ccc}
1 & 1 & 0 \\
1 & 1 & 0 \\
1 & 0 & 0
\end{array}\right), &\quad& $see FIG$.~\ref{FIG_representatives}(12); \\  
\vspace{1mm}
\tilde S_{13}=(1,1,2,3,3) & \quad \rightarrow \quad &\tilde  S_{13}^\mathrm{E}=(1,0,0,\;1,2,3,\;0,0,3) &\quad \rightarrow \quad & \lambda(\tilde S_{13}^\mathrm{E})=\left(\begin{array}{ccc}
0 & 1 & 1 \\
0 & 1 & 0 \\
1 & 1 & 0
\end{array}\right), &\quad& $see FIG$.~\ref{FIG_representatives}(13); \\  
\vspace{1mm}
\tilde S_{14}=(1,1,2,2,3) & \quad \rightarrow \quad &\tilde  S_{14}^\mathrm{E}=(1,0,0,\;1,2,0,\;0,2,3) &\quad \rightarrow \quad & \lambda(\tilde S_{14}^\mathrm{E})=\left(\begin{array}{ccc}
0 & 0 & 1 \\
0 & 1 & 1 \\
1 & 1 & 0
\end{array}\right), &\quad& $see FIG$.~\ref{FIG_representatives}(14); \\  
\vspace{1mm}
\tilde S_{15}=(1,1,2,3,2) & \quad \rightarrow \quad &\tilde  S_{15}^\mathrm{E}=(1,0,0,\;1,2,3,\;0,2,0) &\quad \rightarrow \quad & \lambda(\tilde S_{15}^\mathrm{E})=\left(\begin{array}{ccc}
0 & 1 & 0 \\
0 & 1 & 1 \\
1 & 1 & 0
\end{array}\right), &\quad& $see FIG$.~\ref{FIG_representatives}(15); \\  
\vspace{1mm}
\tilde S_{16}=(1,2,1,2,3) & \quad \rightarrow \quad &\tilde  S_{16}^\mathrm{E}=(1,2,0,\;1,2,3,\;0,0,0) &\quad \rightarrow \quad & \lambda(\tilde S_{16}^\mathrm{E})=\left(\begin{array}{ccc}
0 & 1 & 0 \\
1 & 1 & 0 \\
1 & 1 & 0
\end{array}\right), &\quad& $see FIG$.~\ref{FIG_representatives}(16);  \\
\vspace{1mm}
\tilde S_{17}=(1,1,1,2,3) & \quad \rightarrow \quad &\tilde  S_{17}^\mathrm{E}=(1,0,0,\;1,0,0,\;1,2,3) &\quad \rightarrow \quad & \lambda(\tilde S_{17}^\mathrm{E})=\left(\begin{array}{ccc}
0 & 0 & 1 \\
0 & 0 & 1 \\
1 & 1 & 1
\end{array}\right), &\quad& $see FIG$.~\ref{FIG_representatives}(17); 
\end{array}
\end{equation}

\begin{figure*}
\includegraphics[width=0.9\textwidth]{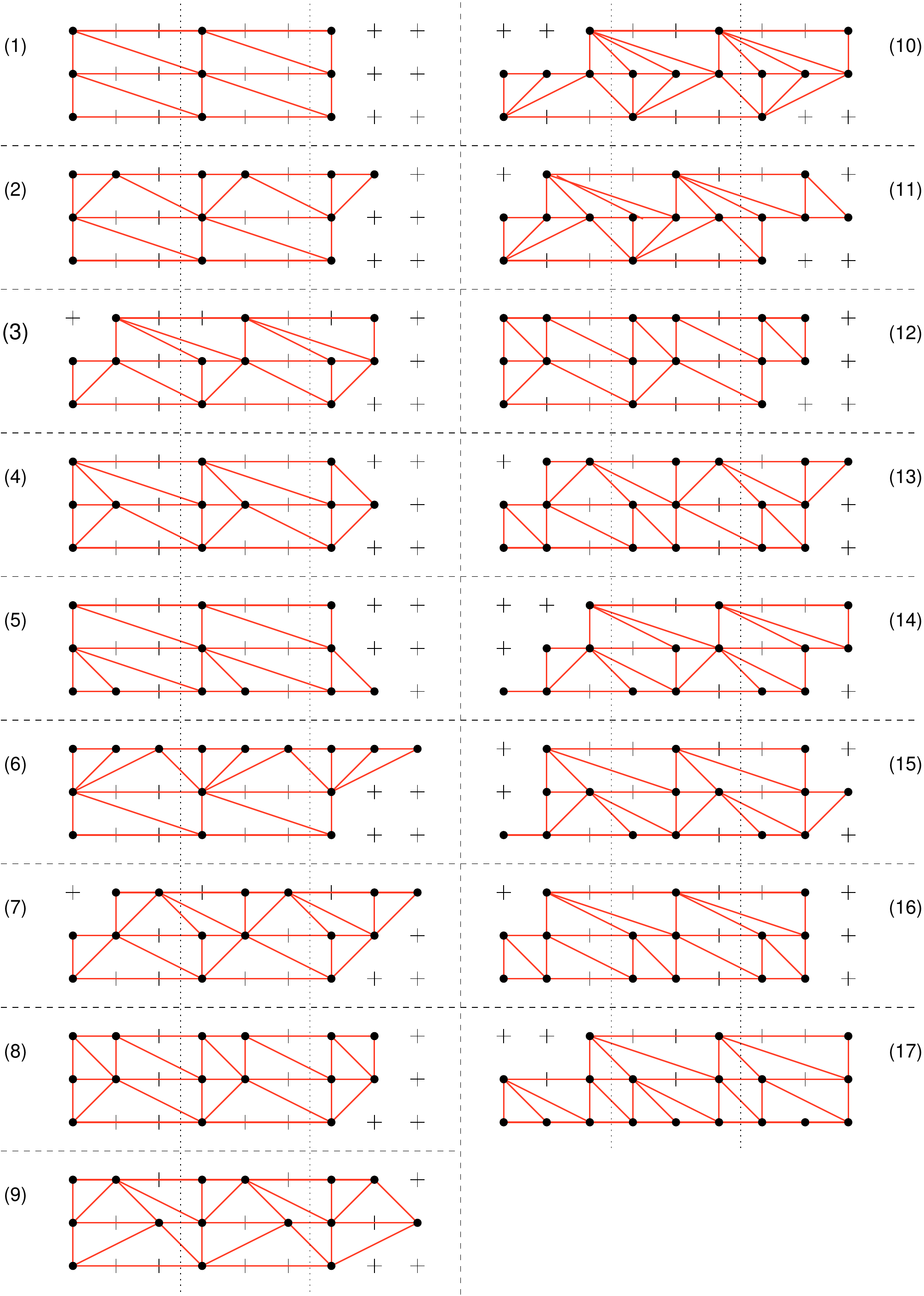}
\caption{The representatives of the equivalence classes for all triangulations up to $5$ forks distributed on $3$ spatial slices are shown. Notice that because of the periodic boundary conditions on the spatial slices, it is advantageous to display three periods (indicated by the dotted vertical lines).}
\label{FIG_representatives} 
\end{figure*}

\section{Degeneracies of the various encodings}
The binary and the integer encoding differ in the size of their configuration spaces. To show this, let us for a moment relax the constraint of only considering connected triangulations to get a rough estimate of the size of the configuration space. For $F\gg N$ we can simplify the minimum lattice size required to encode all triangulations to $N \times F$. Thus the number of triangulations in the binary encoding is growing $\propto (2^N)^F$, while the size of the integer encoding grows  $\propto N^F$. 
Since both encodings are ergodic (in the sense of containing all triangulations with $F$ forks distributed over $N$ spatial slices at least once) the binary encoding is less efficient than the integer encoding. 
The integer encoding has the additional advantage that the degeneracies of the configuration space are related to algebraic symmetry operations acting on the strings (see above).

This can be demonstrated with the help of the following example. Consider the integer string $S=(1,2,3,\;1,2,3,\;1,2,3)$, representing a triangulation with coordination number $6$ everywhere except at the boundaries. Even for this trivial example the mapping to a binary encoding is not unique, if the lattice size is larger then $3\times3$. In Fig.~\ref{FIG_degeneracy} we have depicted $6$ seemingly different encodings of $S$. Fig.~\ref{FIG_degeneracy}(a-d) are simply different binary encodings of $S$, while Fig.~\ref{FIG_degeneracy}(e-f) are mappings of $S$ to $\lambda$ after having applied various symmetry operations on $S$. 

\begin{figure}[h!]
\includegraphics[width=0.8\textwidth]{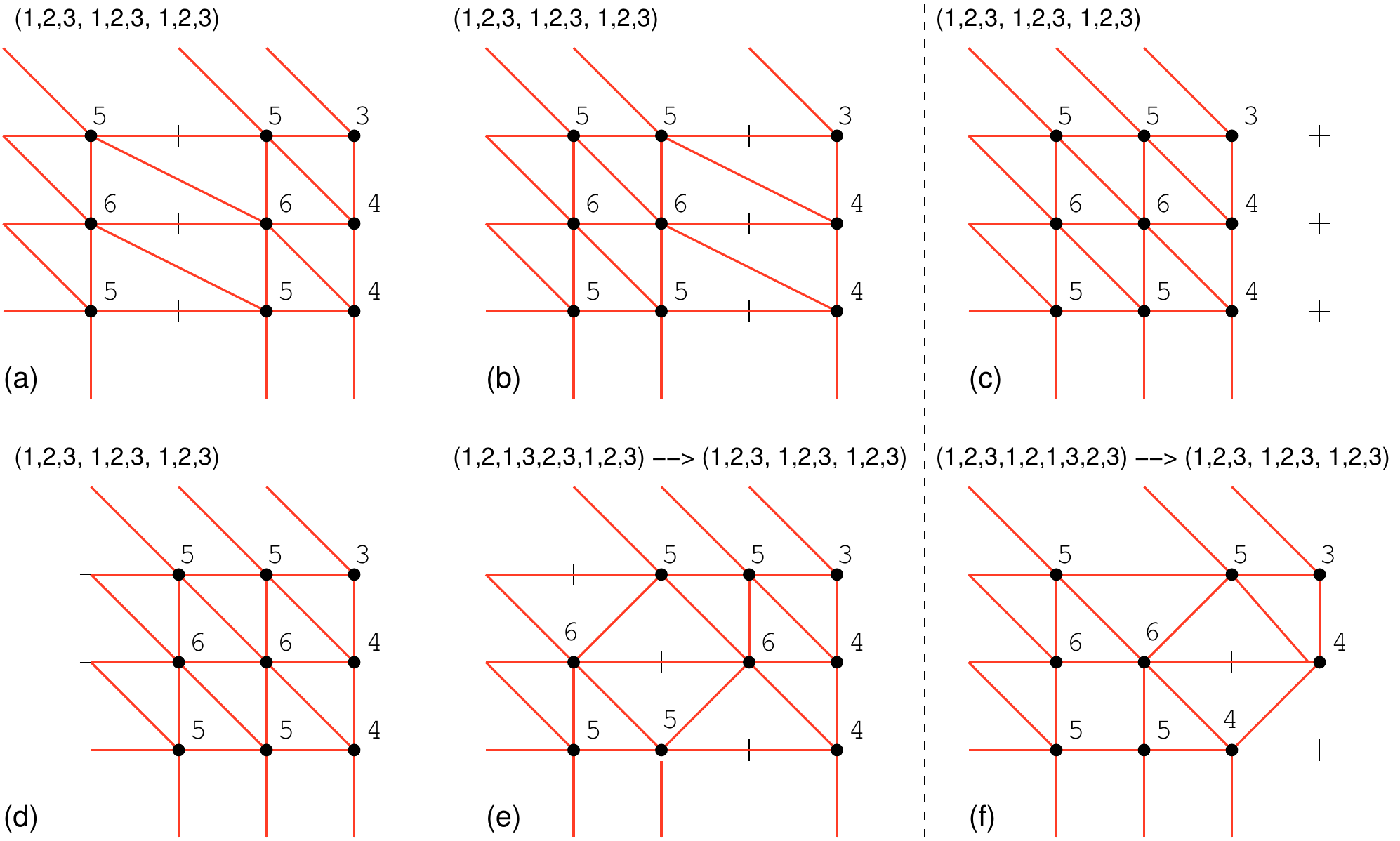}
\caption{\label{FIG_degeneracy} Six different binary encodings of $\bar{S}=(1,2,3,\;1,2,3,\;1,2,3)$ in a $3\times 4$ lattice are represented. The numbers at the vertices are the coordination numbers.}
\end{figure}
\end{widetext} 
\end{document}